\documentclass[12pt]{cernart}
\usepackage{graphicx}
\usepackage{epsfig}
\usepackage{multirow}
\newcommand{\defmath}[2] {\def#1{\ifmmode{#2}\else\mbox{${#2}$}\fi}} 
\newcommand{\defdecay}[2] {\def#1{\ifmmode{#2}\else{${#2}$}\fi}}
\newcommand{\E}[1] {\mathrm{\times 10^{#1}}}

\newcommand{\Br}[1] {\mathrm{\Gamma}({#1})}

\defdecay{\k} {K}
\defdecay{\kz} {K^0}
\defdecay{\kbar} {\overline{K}^{_{\scriptstyle 0}}}
\defdecay{\kone} {K_1}
\defdecay{\ktwo} {K_2}
\defdecay{\ks} {K_S}
\defdecay{\kl} {K_L}
\defdecay{\ksl} {K_{S,L}}
\defdecay{\KL} {\kl}
\defdecay{\KS} {\ks}

\defdecay{\pipi} {\pi\pi}
\defdecay{\twopi} {2\pi}
\defdecay{\kpipi} {\k \rightarrow \pipi}
\defdecay{\ktwopi} {\ks \rightarrow \twopi}
\defdecay{\kspipi} {\ks \rightarrow \pipi}
\defdecay{\kstwopi} {\kl \rightarrow \twopi}
\defdecay{\klpipi} {\kl \rightarrow \pipi}
\defdecay{\kltwopi} {\kl \rightarrow \twopi}
\defdecay{\pipipi} {\pi\pi\pi}
\defdecay{\threepi} {3\pi}

\defdecay{\pio} {\pi^0}
\defdecay{\twopio} {2 \pi^0}
\defdecay{\piopio} {\pi^0 \pi^0}
\defdecay{\pipin} {\piopio}
\defdecay{\ktwopio} {\k \rightarrow \twopio}
\defdecay{\kpiopio} {\k \rightarrow \piopio}
\defdecay{\kltwopio} {\kl \rightarrow \twopio}
\defdecay{\klpiopio} {\kl \rightarrow \piopio}
\defdecay{\kstwopio} {\ks \rightarrow \twopio}
\defdecay{\kspiopio} {\ks \rightarrow \piopio}

\defdecay{\pip} {\pi^+}
\defdecay{\pim} {\pi^-}
\defdecay{\twopic} {\pi^+ \pi^-}
\defdecay{\pipic} {\twopic}
\defdecay{\ktwopic} {\k \rightarrow \twopic}
\defdecay{\kltwopic} {\kl \rightarrow \twopic}
\defdecay{\kstwopic} {\ks \rightarrow \twopic}

\defdecay{\piod} {\pi_{D}^0}
\defdecay{\piopiod} {\pi^0 \pi_{D}^0}
\defdecay{\eeg} {e^+e^-\gamma}
\defdecay{\pioeeg} {\pi^0 \rightarrow \eeg}
\defdecay{\kpiopiod} {\k \rightarrow \piopiod}

\defdecay{\kethree} {\mathrm{K_{e3}}}
\defdecay{\pienu} {\pi e \nu}
\defdecay{\klethree} {\kl \rightarrow \pienu}
\defdecay{\kmuthree} {\mathrm{K_{\mu 3}}}
\defdecay{\pimunu} {\pi \mu \nu}
\defdecay{\klmuthree} {\kl \rightarrow \pimunu}
\defdecay{\threepio} {3\pi^0}
\defdecay{\klthreepio} {\kl \rightarrow \threepio}

\defdecay{\Lam}{\Lambda}
\defdecay{\Lambar}{\bar{\Lambda}}
\defdecay{\etagg}{\eta \rightarrow \gamma \gamma}
\defdecay{\etathreepio} {\eta \rightarrow \threepio}
\defdecay{\piogg} {\pi^0 \rightarrow \gamma \gamma}

\defmath{\dvertex}{d_{vertex}}
\defmath{\mgg}{m_{\gamma\gamma}}
\defmath{\chisq}{\chi^2}
\defmath{\mone}{m_1}
\defmath{\mtwo}{m_2}

\defmath{\pk}{p_K}
\defmath{\pt}{p_T}
\defmath{\ptp}{{p_T}'}
\defmath{\ptpsq}{{p_T}'^2}
\defmath{\mpp}{m_{\pi\pi}}

\defmath{\asl} {\alpha_{SL}}
\defmath{\asloo} {\alpha^{00}_{SL}}
\defmath{\aslpm} {\alpha^{+-}_{SL}}
\defmath{\Dasl} {\Delta \alpha_{SL}}

\defmath{\als} {\alpha_{LS}}
\defmath{\alsoo} {\alpha^{00}_{LS}}
\defmath{\alspm} {\alpha^{+-}_{LS}}
\defmath{\Dals} {\Delta \alpha_{LS}}

\defmath{\btag} {\beta_{tag}}
\defmath{\btagoo} {\beta^{00}_{tag}}
\defmath{\btagpm} {\beta^{+-}_{tag}}
\defmath{\Dbtag} {\Delta \beta_{tag}}

\defmath{\wpm} {W^{+-}}
\defmath{\woo} {W^{00}}
\defmath{\wooo} {W^{000}}
\defmath{\Dw} {\Delta W}

\defmath{\wt}{W(\tau)}
\defmath{\Dp}{\mathrm{D_p}}

\defmath{\etas}{\eta_S}
\defmath{\etal}{\eta_L}
\defmath{\etasl}{\eta_{S,L}}
\defmath{\lamc}{\lambda^{+-}}
\defmath{\lamn}{\lambda^{00}}
\defmath{\DRint}{(\DR)_{\mbox{\scriptsize intensity}}}
\defmath{\DRgeom}{(\DR)_{\mbox{\scriptsize geometry}}}

\defmath{\R}{R}
\defmath{\DR}{\Delta \R}
\defmath{\epp}{\varepsilon^{\prime}}
\defmath{\vep}{\varepsilon}
\defmath{\epe}{\epp/\vep}
\defmath{\Ree}{\mathrm{Re}(\epe)}
\defmath{\epm}{\eta_{+-}}
\defmath{\eoo}{\eta_{00}}

\defmath{\mum}{\mu\mathrm{m}}
\defmath{\mus}{\mu\mathrm{s}}
\defmath{\degrees}{^{\circ}}
\defmath{\taus}{\tau_S}
\defmath{\taul}{\tau_L}

\defmath{\about}{\sim}
\defmath{\eop}{E/p}
\defmath{\Qx} {Q_x}
\defmath{\twotrack}{2track}
\defmath{\etot}{E_{tot}}
\defmath{\mk}{m_K}
\defmath{\stat}{\mbox{stat}}
\defmath{\syst}{\mbox{syst}}

\begin{document}
\begin{titlepage}
\docnum{CERN--EP/2002--061}
\date{25 July 2002}
\title{
 \boldmath A precision measurement of direct CP violation \\
  in the decay of neutral kaons into two pions\\ 
}
\begin{Authlist}

The NA48 Collaboration \\

\begin{center}
J.R.~Batley,
R.S.~Dosanjh,
T.J.~Gershon\footnotemark[1],
G.E.~Kalmus,
C.~Lazzeroni,
D.J.~Munday,
E.~Olaiya,
M.~Patel,
M.A.~Parker,
T.O.~White,
S.A.~Wotton \\
{\em Cavendish Laboratory, University of Cambridge, Cambridge, CB3 0HE, U.K.\footnotemark[2].} \\[0.2cm] 
R.~Arcidiacono,
G.~Barr,
G.~Bocquet,
A.~Ceccucci,
T.~Cuhadar-D\"{o}nszelmann,
D.~Cundy,
N.~Doble,
V.~Falaleev,
L.~Gatignon,
A.~Gonidec,
B.~Gorini,
P.~Grafstr\"om,
W.~Kubischta,
I.~Mikulec\footnotemark[3],
A.~Norton,
S.~Palestini,
B.~Panzer-Steindel,
D.~Schinzel,
H.~Wahl \\
{\em CERN, CH-1211 Gen\`eve 23, Switzerland.} \\[0.2cm] 
C.~Cheshkov,
P.~Hristov,
V.~Kekelidze,
D.~Madigojine,
N.~Molokanova,
Yu.~Potrebenikov,
A.~Zinchenko \\
{\em Joint Institute for Nuclear Research, Dubna, Russian    Federation.} \\[0.2cm] 
P.~Rubin\footnotemark[4],
R.~Sacco,
A.~Walker \\
{\em Department of Physics and Astronomy, University of    Edinburgh, Edinburgh,    EH9 3JZ, U.K.\footnotemark[2].} \\[0.2cm] 
D.~Bettoni,
R.~Calabrese,
P.~Dalpiaz,
J.~Duclos,
P.L.~Frabetti\footnotemark[5],
A.~Gianoli,
M.~Martini,
L.~Masetti,
F.~Petrucci,
M.~Savri\'e,
M.~Scarpa \\
{\em Dipartimento di Fisica dell'Universit\`a e Sezione    dell'INFN di Ferrara, I-44100 Ferrara, Italy.} \\[0.2cm] 
A.~Bizzeti\footnotemark[6],
M.~Calvetti,
G.~Collazuol,
E.~Iacopini,
M.~Lenti,
F.~Martelli\footnotemark[7],
G.~Ruggiero,
M.~Veltri\footnotemark[7] \\
{\em Dipartimento di Fisica dell'Universit\`a e Sezione dell'INFN di Firenze, I-50125 Firenze, Italy.} \\[0.2cm] 
D.~Coward,
M.~Eppard,
A.~Hirstius,
K.~Holtz,
K.~Kleinknecht,
U.~Koch,
L.~K\"opke,
P.~Lopes~da~Silva, 
P.~Marouelli,
I.~Mestvirishvili,
C.~Morales,
I.~Pellmann,
A.~Peters,
B.~Renk,
S.A.~Schmidt,
V.~Sch\"{o}nharting,
R.~Wanke,
A.~Winhart\\
{\em Institut f\"ur Physik, Universit\"at Mainz, D-55099    Mainz, Germany\footnotemark[8].} \\[0.2cm] 
J.C.~Chollet,
L.~Fayard,
G.~Graziani,
L.~Iconomidou-Fayard,
G.~Unal,
I.~Wingerter-Seez \\
{\em Laboratoire de l'Acc\'el\'erateur Lin\'eaire,  IN2P3-CNRS,Universit\'e de Paris-Sud, 91898 Orsay, France\footnotemark[9].} \\[0.2cm] 
G.~Anzivino,
P.~Cenci,
E.~Imbergamo,
G.~Lamanna,
P.~Lubrano,
A.~Mestvirishvili,
A.~Nappi,
M.~Pepe,
M.~Piccini,
M.~Valdata-Nappi \\
{\em Dipartimento di Fisica dell'Universit\`a e Sezione    dell'INFN di Perugia, I-06100 Perugia, Italy.} \\[0.2cm] 
R.~Casali,
C.~Cerri,
M.~Cirilli\footnotemark[10],
F.~Costantini,
R.~Fantechi,
L.~Fiorini,
S.~Giudici,
I.~Mannelli, 
G.~Pierazzini,
M.~Sozzi \\
{\em Dipartimento di Fisica, Scuola Normale Superiore e Sezione dell'INFN di Pisa, I-56100 Pisa, Italy.} \\[0.2cm]   
J.B.~Cheze,
M.~De Beer,
P.~Debu,
F.~Derue,
A.~Formica,
G.~Gouge,
G.~Marel,
E.~Mazzucato,
B.~Peyaud,
R.~Turlay,
B.~Vallage \\
{\em DSM/DAPNIA - CEA Saclay, F-91191 Gif-sur-Yvette, France.} \\[0.2cm] 
M.~Holder,
A.~Maier,
M.~Ziolkowski \\
{\em Fachbereich Physik, Universit\"at Siegen, D-57068 Siegen, Germany\footnotemark[11].} \\[0.2cm] 
C.~Biino,
N.~Cartiglia,
M.~Clemencic,
F.~Marchetto, 
E.~Menichetti,
N.~Pastrone \\
{\em Dipartimento di Fisica Sperimentale dell'Universit\`a e    Sezione dell'INFN di Torino,  I-10125 Torino, Italy.} \\[0.2cm] 
J.~Nassalski,
E.~Rondio,
W.~Wislicki,
S.~Wronka \\
{\em Soltan Institute for Nuclear Studies, Laboratory for High    Energy Physics,  PL-00-681 Warsaw, Poland\footnotemark[12].} \\[0.2cm] 
H.~Dibon,
M.~Jeitler,
M.~Markytan,
G.~Neuhofer,
M.~Pernicka,
A.~Taurok,
L.~Widhalm \\
{\em \"Osterreichische Akademie der Wissenschaften, Institut  f\"ur Hochenergiephysik,  A-1050 Wien, Austria\footnotemark[13].} 
\end{center}
\end{Authlist}
\footnotetext[1]{Present address: High Energy Accelerator Research
               Organization (KEK), Tsukuba, Ibaraki, 305-0801, Japan.}
\footnotetext[2]{ Funded by the U.K.    Particle Physics and Astronomy Research Council.}
\footnotetext[3]{ On leave from \"Osterreichische Akademie der Wissenschaften, Institut  f\"ur Hochenergiephysik,  A-1050 Wien, Austria.}
\footnotetext[4]{Permanent address: Department of Physics,
University of Richmond, VA 27313, USA}
\footnotetext[5]{Dipartimento di Fisica e INFN Bologna, viale
  Berti-Pichat 6/2, I-40127 Bologna, Italy.}
\footnotetext[6]{ Dipartimento di Fisica
dell'Universita' di Modena e Reggio Emilia, via G. Campi 213/A
I-41100, Modena, Italy.}
\footnotetext[7]{Instituto di Fisica Universit\'a di Urbino}
\footnotetext[8]{ Funded by the German Federal Minister for    Research and Technology (BMBF) under contract 7MZ18P(4)-TP2.}
\footnotetext[9]{ Funded by Institut National de Physique des
  Particules et de Physique Nucl\'eaire (IN2P3), France}
\footnotetext[10]{Present address: Dipartimento di Fisica
  dell'Universit\'a di Roma ``La Sapienza'' e Sezione INFN di Roma,
  I-00185 Roma, Italy.}
\footnotetext[11]{ Funded by the German Federal Minister for Research and Technology (BMBF) under contract 056SI74.}
\footnotetext[12]{    Supported by the Committee for Scientific Research grants 5P03B10120, 2P03B11719 and SPUB-M/CERN/P03/DZ210/2000 and using computing resources of the Interdisciplinary Center for    Mathematical and    Computational Modelling of the University of Warsaw.} 
\footnotetext[13]{    Funded by the Austrian Ministry of Education,
  Science and Culture under contract GZ 616.360/2-IV GZ
  616.363/2-VIII, and by the Fund for Promotion of Scientific Research
  in Austria (FWF) under contract P08929-PHY.}
\abstract{
The direct CP violation parameter 
Re($\epsilon'/\epsilon)$ has been measured from the decay
rates of neutral kaons into two pions using the NA48 detector at the CERN
SPS. 
The 2001 running period was devoted to collecting additional data under
varied conditions compared to earlier years (1997-99).
The new data yield the result: Re($\epsilon'/\epsilon)$=$(13.7\pm3.1)\times10^{-4}$.
Combining this result with that published from the 1997, 98 and 99 data,
an overall value of 
Re($\epsilon'/\epsilon)$=$(14.7\pm2.2)\times10^{-4}$ is obtained from
the NA48 experiment.
} 
\submitted{(To be published in Physics Letters B)}

\maketitle

\end{titlepage}
\renewcommand{\thefootnote}{\arabic{footnote}}

\section{Introduction}
\label{sec:intro}
CP violation was discovered 38 years ago in the decays of neutral 
kaons~\cite{disco}. Recently CP violation in the 
$B^0$ mesons has also been observed~\cite{sin2b}; nevertheless neutral kaons remain a 
privileged system for the study of the phenomenon.

CP conservation would imply that the \ks\ and \kl\ particles are pure
CP-eigenstates and that \kl\ decay only into CP=$-$1 and \ks\ into
CP=+1 final states. 
The observed signal of the forbidden \kltwopi\ decays (CP=$+$1) indicates that 
CP is not a conserved symmetry.

CP violation can occur via the mixing of CP eigenstates,
called {\em indirect\/} CP violation,
represented by the parameter $\epsilon$. 
CP violation can also occur in the decay process
through the interference of amplitudes with different isospins. 
This is represented by the parameter $\epsilon'$
and is called {\em direct\/} CP violation. 

In the Standard Model of electro-weak interaction, 
CP violation is naturally accommodated 
by an irreducible complex phase in the quark mixing-matrix~\cite{kob}. 
Current theoretical  predictions of $\epsilon'/\epsilon$
range \about$-10\E{-4}$ to \about$+30\E{-4}$~\cite{theory}.

Experimentally, it is convenient to measure 
the double ratio \R\ of decay widths, 
which is related  to the ratio $\epsilon'/\epsilon$ as follows:
\begin{equation}
\label{eq:doub}
\R = \frac{ \Br{\klpiopio} }{ \Br{\kspiopio} } / \frac{ \Br{\kltwopic} }{ \Br{\kstwopic} }
  \approx 1 - 6 \times \mathrm{Re}(\epsilon'/\epsilon)
\end{equation}

In 1993, two experiments  published their final results: 
NA31~\cite{na31} measured Re($\epsilon'/\epsilon)=(23.0\pm6.5)\E{-4}$, 
and the result of E731~\cite{e731} was 
Re($\epsilon'/\epsilon)=(7.4\pm5.9)\E{-4}$. 
Recently, two experiments  
announced results from samples of their total statistics. 
NA48 published a result of Re($\epsilon'/\epsilon)=(15.3\pm2.6)\E{-4}$,
using data collected in 1997~\cite{na48_97}, 1998 and 99~\cite{na48_99}, 
and KTeV presented a preliminary result 
of  Re($\epsilon'/\epsilon)=(20.7\pm2.8)\E{-4}$ \cite{ktev}
on data
accumulated in 1996~\cite{ktev_publ} and 97.
These  observations confirmed the existence of a 
direct CP-violation component.

This paper reports a measurement of Re($\epsilon'/\epsilon)$ performed using
the 2001 data sample, recorded in 
somewhat different experimental conditions by the NA48 experiment.

After the 1999 data-taking period, the drift chambers of the experiment
were damaged by the implosion of the beam tube. The data
taking in 2001 took place with rebuilt drift chambers. Thanks
to the possibility of a better SPS duty cycle, increased
by a factor 1.8 with respect to the 1998-99 running period, the
data could be taken at a 30\% lower beam intensity,
allowing the insensitivity of the result to intensity-related
effects to be checked, 
and the statistics for the final
$\epsilon'/\epsilon$ measurement by NA48 to be completed.
The statistics accumulated during
the 93 days of the 2001 data-taking period is roughly half of the
total statistics accumulated in the 263 days 
of the 1998 and 99 periods.

Details of the apparatus and of the data 
analysis can be found in~\cite{na48_99}, here 
only the differences with respect to the 1998-99 data-taking will be stressed.

\section{The method}
\label{sec:method}

Re($\epsilon'/\epsilon$) is derived from the double ratio R.
The experiment is designed to exploit cancellations of systematic effects
contributing symmetrically
to different components of the double ratio.

The four decay modes are collected simultaneously, which minimises the
sensitivity of the measurement to accidental activity and to variations
in beam intensity and detection efficiency. 
In the analysis $K_S$ events are further weighted 
by the $K_L$/$K_S$ intensity ratio to 
eliminate the small and slow variations of the $K_L$ and $K_S$ beam
intensities.
To maintain the simultaneous data-taking of $\pi^0\pi^0$ and
$\pi^+\pi^-$ decays, dead-time conditions affecting one mode are recorded
and applied offline in all modes.

$K_L$ and $K_S$ decays are provided by two nearly-collinear beams with
similar momentum spectra, converging to the centre of the main detector.
The same decay region is used for all modes. In order to minimise
the acceptance correction
due to the difference in mean decay lengths, $K_L$
decays are weighted as a function of their proper lifetime,
such that the $K_L$ decay distribution becomes similar to that of
$K_S$. In this way, the accuracy of the result does not rely on a detailed
Monte Carlo simulation of the experiment and only small remaining differences
in beam divergences and geometries need to be corrected using Monte Carlo
simulation. To be insensitive to residual differences in the beam
momentum spectra, the analysis is performed in bins of kaon energy.

$K_S$ decays are distinguished from $K_L$ decays by a coincidence between
the decay time and the registered times of the protons producing the $K_S$ beam.
As the same method is used for $\pi^+\pi^-$ and
$\pi^0\pi^0$ decays, the double ratio is sensitive only to differences
in misidentification probabilities between the two decay modes and not
to their absolute values.

Finally, high-resolution detectors are used to detect the $\pi^+\pi^-$ and
$\pi^0\pi^0$ final states in order to minimise residual backgrounds which
do not cancel in the double ratio.

\vspace{0.5cm}
\section{Beams and detectors}
\label{sec:beamdet}

  \subsection{\boldmath Beams}
  \label{sec:beam}
The \KL\ and \KS\ beams~\cite{beampaper} 
are produced in  two different
targets by protons from the same CERN SPS 
beam.
In the 2001 run the SPS had a cycle time of 16.8~s with a spill length of 5.2~s 
and a proton momentum of 400~GeV/$c$ \footnote{The cycle time was 14.4~s, the spill 
length 2.38~s and the proton momentum 450~GeV/$c$ in the 
1998 and 99 runs.
The effective spill length, given by the remaining time structures in the
beam is $\approx$3.6~s in 2001, 
compared to $\approx$1.7~s for the 1998-99 data.}.
Since the \ks\ and \kl\ beams are produced concurrently, 
the \ks/\kl\ ratio is maintained stable 
to within $\pm10\%$. 

\par
The primary, high-flux proton beam ($\about2.4\E{12}$~protons per pulse) 
impinges on the \kl\ target (a 400~mm long, 2~mm diameter rod of beryllium), 
with an incidence angle of 2.4~mrad relative to the \kl\ beam axis.
The charged component of the outgoing  particles
is swept away by bending magnets. 
The neutral beam passes through three stages of collimation and
the fiducial region starts at the exit of the ``final'' collimator,
126~m downstream of the target. 
At this point, this neutral beam is dominated by long-lived kaons,
neutrons and photons;
only small fractions of the most energetic short-lived 
components (\ks\ and \Lam) survive.

\par
The protons not interacting in the \kl\ target 
are directed onto a mechanically bent mono-crystal of silicon~\cite{crystal}.
A small fraction of protons satisfies the conditions for channelling and
is bent to produce a collimated beam of \about5$\E{7}$ protons per pulse,
which is then deflected back onto the \kl\ beam axis and finally directed 
to the \ks\ target (of similar dimensions as the \kl\ target) 
located 72~mm above the \kl\ beam axis. 
A combination of a sweeping magnet and collimator selects a neutral beam at 
4.2~mrad to the incoming protons. 
The decay spectrum of kaons at the exit of the collimator 
(6 m downstream of the target) is similar to that in the
\kl\ beam, with an average energy of 110~GeV \footnote{Despite the different 
proton momentum in the 1998 and 99 runs, the $K_L$ and $K_S$ spectra remain similar,
depending only on the choice of production angles to compensate for the length
of the $K_S$ collimator.}. 
Two-pion decays from this beam come almost exclusively from $K_S$ decays.
\par
The tagging station (Tagger) is located on the path of the \ks\ proton beam after
the bent crystal. It consists of two ladders of 12 scintillator strips each, 
covering the beam horizontally and vertically~\cite{tagger}. 
A proton crosses at least two scintillators, one horizontal and one vertical. 
The reconstructed time per counter has a resolution of \about 140~ps,
and two pulses 4--5~ns apart can be resolved.
\par
The beginning of the \ks\ decay region is sharply defined by an
anti-counter (AKS), located at the exit of the \ks\ collimator~\cite{akspap}.
It is composed of a photon converter followed by three scintillator counters.
Its main purpose is to veto all upstream decays from the $K_S$ beam.
\par
The decay region is contained in an evacuated
($< 3 \times 10^{-5}$~mbar) 90~m long tank with a
0.9~mm (0.003~radiation length) thick polyamide (Kevlar)
composite window at the end. The neutral beam continues in
a 16~cm diameter evacuated tube to the beam dump, downstream
of all detector elements.

\subsection{\boldmath Detectors}
\label{sec:detector}

Charged particles from decays are measured by a magnetic spectrometer~\cite{chambers}
composed of four drift chambers with a dipole magnet (inducing a 
transverse momentum-kick of 265~MeV/$c$ in the horizontal plane) 
between the second and third chambers.
These chambers and their interconnecting beam tube are aligned 
along the bisector between the converging \ks\ and \kl\ beam axes.
Each chamber is comprised of eight planes of sense wires, two horizontal,
two vertical and two along each of the 45$^\circ$ directions.
In the third chamber,
only the horizontal and vertical planes are instrumented.
The average efficiency per plane is $99.5\%$, with a radial uniformity
better than $\pm0.2\%$. The space point resolution is $\approx95$ $\mu$m.
The momentum resolution is
$\sigma(p)/p = 0.48\% \oplus 0.009\%\times p$, where the momentum
$p$ is in GeV/$c$. These performance figures are similar to those
obtained previously \cite{na48_99}.

\par
The magnetic spectrometer is followed by a scintillator hodoscope, 
composed of two planes segmented in horizontal and vertical strips. 
Fast logic combines the strip signals (arranged in four quadrants)
for use in the first level of the $\pi^+\pi^-$ trigger. 
\par
A liquid Krypton calorimeter (LKr) is used to reconstruct 
K$\rightarrow 2\pi^{0}$ decays. 
It is a quasi-homoge\-neous detector with an active volume of 
\about 10~m$^{3}$ of liquid Krypton. Cu-Be-Co ribbon electrodes
of $40~\mum \times 18~\mathrm{mm} \times 125~\mathrm{cm}$ define 
\about 13000 cells (each with 2~cm $\times$ 2~cm cross-section)
in a structure of longitudinal projective towers 
pointing to the centre of the decay region~\cite{caloref}. 
The calorimeter is 27 radiation lengths long
and fully contains electro-magnetic showers with energies up to 100 GeV. 
The energy resolution of the calorimeter is
$\sigma(E)/E = (3.2\pm0.2)\%/\sqrt{E}
            \oplus (9\pm1)\% /E
            \oplus (0.42\pm0.05)\% $
with $E$ in GeV~\cite{calor2000}.
The LKr calorimeter also is used,
together with an iron-scintillator calorimeter,
to measure the total deposited energy for triggering purposes.
\par
Finally, at the end of the beam 
line, muon counters are used to identify \kl$\rightarrow \pi\mu\nu$ (K$_{\mu3}$) decays. 
\par
Two beam counters are used to measure the intensity of the beams: 
one is located at the extreme end of the \kl\ beam line (\kl\ monitor) and 
the other (\ks\ monitor) views the \ks\ target station.
For the 2001 data taking, another \kl\ monitor with a higher
counting rate was added and a \ks\ monitor near the
tagging station was installed. These allow
better measurements of the beam structures to be made down to a time
scale of $\approx$ 200~ns.
 
\subsection{\boldmath Triggers}
\label{sec:tridaq}

The rate of particles reaching the detector is around  400~kHz. The
trigger is designed to reduce this rate to less than 10~kHz,
with minimal loss from dead time and inefficiencies. 
A part of the read-out rate is reserved for
redundant low-bias triggers that collect data used for the direct 
determination of the trigger inefficiencies. 

Triggers initiated by the beam monitors
are used to record the accidental activity,
with rates proportional to \KL\ and \KS\ decay rates. 
Beam monitor signals are down-scaled and delayed by 69~\mus\, 
corresponding to the periodicity of the slow proton extraction 
(3 SPS revolutions).

\subsubsection{Trigger for \pipin\ decays}
\label{sec:trign}
The trigger for $\pipin$ decays~\cite{nut} operates digitally on the 
analogue sums of signals from $2\times8$ cells 
(in both horizontal and vertical orientations) of the LKr calorimeter.
These sums are converted into kinematic quantities
by a ``look-up table'' system.

The trigger requires
an electro-magnetic energy deposit greater than 50~GeV,
a centre of energy (distance between the extrapolated kaon impact point 
at the calorimeter plane and the beam axis) smaller than 25~cm, 
and a decay vertex less than 5 \ks\ lifetimes (\taus)
from the beginning of the decay volume.
Requesting 
less than 6 peaks within 9 ns in both projections 
helps to reject background from \klthreepio (this condition
is released if accidental activity is detected close in time).

A trigger for \threepio\ decays, given by the down-scaled \pipin\ trigger
without the peak condition, is used for tagging studies.

\subsubsection{Trigger for \pipic\ decays}
\label{sec:trigc}

The $\pipic$ decays are triggered with a two-level trigger system. 
At the first level, 
the rate is reduced to 100~kHz by a coincidence of three fast signals:
opposite quadrant coincidence in the scintillator
hodoscope (\Qx), hit multiplicity in the first drift chamber integrated over 200~ns, 
and the total calorimetric energy (\etot, with a threshold of 35~GeV).

The second level of the \pipic\ trigger~\cite{mbx}, consisting of 
hardware coordinate builders 
and a farm of asynchronous microprocessors,
reconstructs tracks using data from the drift chambers.
Triggers are accepted if the tracks converge to within 5~cm, 
their opening angle is smaller than 15~mrad, 
the reconstructed proper decay time is smaller than 4.5 \taus\
and the reconstructed $\pi\pi$ mass is greater than 0.95 \mk. 

\section{Event reconstruction and selection}
\label{sec:reco}

\subsection{\boldmath $\pipin$}
\label{sec:recon}

 K $\rightarrow$ $\pi^0\pi^0$ decays are selected using only data from
the LKr calorimeter. The reconstruction of photon showers and the details
of the small corrections applied to the energy and position measurements can
be found in~\cite{na48_99} and~\cite{eta_mass}.
 Photon showers in the energy range $3-100$ GeV are used. Fiducial cuts
are applied to ensure that the photon energies are well measured: the
shower position should be more than 15~cm away from 
the axis of the beam tube,
more than 11~cm away from the outer edges of the calorimeter and more
than 2~cm away from a defective calorimeter channel ($\approx$ 0.4\% of the
channels).
 $\pi^0\pi^0$ decays are selected by requiring four showers 
which are reconstructed within $\pm$~5~ns of their average time
and fulfill the cuts above.
The minimum distance between photons is required to be more than 10~cm.
 To reduce the background from $K_L \rightarrow 3\pi^0$ decays, events  
are rejected which have
an additional cluster of energy above 1.5~GeV and within $\pm$~3~ns of the
time of the $\pi^0\pi^0$ candidate.

 From the measured photon energies $E_i$ and impact point positions on
the calorimeter $x_i$,$y_i$, the distance $D$ from the decay vertex 
to the calorimeter is computed as follows, assuming that
the invariant mass of the four showers is the kaon mass ($m_K$):
\begin{equation}
D = \sqrt{\Sigma_i \Sigma_{j>i} E_i E_j ( (x_i - x_j)^2 + (y_i - y_j)^2)}/m_K .
\end{equation}
 The average resolution on the decay vertex position is about 55~cm, and the resolution
on the kaon energy is $\approx$ 0.5\%.

 The invariant masses $m_1$ and $m_2$ of the two photon pairs are
computed using $D$ and compared to the nominal $\pi^0$ mass 
($m_{\pi^0}$). For this, the following $\chi^2$ variable is constructed:
\begin{equation}
 \chi^2 =
\left[ \frac{(m_1+m_2)/2-m_{\pi^0}}{\sigma _+} \right ]^2 \; + \;
\left[ \frac{(m_1-m_2)/2}{\sigma_-} \right ]^2 .
\end{equation}
The mass combinations $m_1+m_2$ and $m_1-m_2$ are
to good approximation uncorrelated. $\sigma_+$ and $\sigma_-$ are
the resolutions of $(m_1+m_2)/2$ and $(m_1-m_2)/2$ parameterised from the
data as a function of the lowest photon energy. Typical values of
$\sigma_+$ and $\sigma_-$ are 0.4 and 0.8~MeV/$c^2$. Out of the three possible
photon pairings, the one with the lowest $\chi^2$ value is kept.
 To select good $\pi^0\pi^0$ candidates and reject the residual
background from $K_L \rightarrow 3\pi^0$ decays, the cut
$\chi^2 < 13.5$ is applied. 

 The event time is computed by combining eight time estimators from the
two most energetic cells of each cluster. An average 
resolution of 220 ps is thereby obtained.

\subsection{\boldmath $\pipic$}
\label{sec:recoc}

The \pipic\ events are reconstructed from tracks 
using hits in the drift chambers of the spectrometer;
the track momenta are calculated 
using the measured magnetic field map and alignment constants. 

A vertex position is calculated
for each pair of tracks with opposite charge
after correcting for the small residual
magnetic field due to the magnetisation of the vacuum tank 
($\sim 2\E{-3}$~Tm). 
The average resolution on the longitudinal vertex position is
about 50~cm, 
whereas the transverse position resolution is around 2~mm.
Since the beams are separated vertically by about 6~cm in
the decay region, a clean identification of $K_S$ and $K_L$
decays is possible using the reconstructed vertex position.

 Only tracks with momenta greater than 10~GeV/$c$
and not closer than 12~cm to the centre of each chamber 
are accepted. 
The separation of the two tracks at their closest approach 
is required to be less than 3~cm. 
The track positions, extrapolated downstream, 
are required to be 
within the acceptance of the LKr calorimeter 
and of the muon veto system, 
in order to ensure proper electron and muon identification.

 The kaon energy is computed from the opening angle $\theta$
of the two tracks upstream of the magnet and from
the ratio of their momenta $p_1$ and $p_2$, assuming a
$\ktwopic$ decay:
\begin{equation}
\label{Eq:eangle}
E_K = \sqrt{\frac{\mathcal{R}}{\theta^2}
        ( m_K^2 - \mathcal{R} m_{\pi}^2 )}
\,\,\,\,\,{\rm where}\,\,\,\,\,
\mathcal{R} = \frac{p_1}{p_2} + \frac{p_2}{p_1} + 2
\end{equation}
This measurement of the kaon energy is independent
of the absolute magnetic field and relies mostly on the knowledge of the
geometry of the detector.

 A variable $\mathcal{A}$ related to the decay orientation in the kaon
rest frame is defined 
as $\mathcal{A} = |p_1 - p_2|/(p_1 + p_2)$. A cut
is applied to $\mathcal{A}$ ($\mathcal{A}<min(0.62,1.08-0.0052\times E_K)$,
where $E_K$ is in GeV),
to remove
asymmetric decays in which one of the tracks could be close to the beam
tube where the Monte Carlo modelling is more critical. This
cut also removes $\Lambda \rightarrow p\pi^-$ decays.

 To reject background from semileptonic $K_L$ decays, 
events with tracks consistent
with being either an electron or a muon are rejected. To identify
electrons, the ratio $E/p$ of the energy of the matching cluster in the
LKr calorimeter to the track momentum is computed. Pion candidate tracks
are required to satisfy $E/p<0.8$. Tracks are identified as muons if hits
are found in time in the muon counters near the extrapolated track impact
point.

 For good $\pi^+\pi^-$ events, the reconstructed mass $m_{\pi\pi}$ should
be consistent with the kaon mass. 
The resolution on the invariant mass $\sigma_m$ is
typically 2.5~MeV/$c^2$. An energy-dependent cut at $\pm3\sigma_m$
is applied. A further reduction of background from semileptonic decays is
achieved with a cut based on the transverse momentum of the kaon. To
define a selection which is as symmetric as
possible between $K_L$ and $K_S$ decays,
the variable \ptp\ is used, defined as the component of the kaon momentum
orthogonal to the line joining the production target and the point
where the kaon trajectory crosses the plane of the first drift chamber.
To select $\pi^+\pi^-$ candidates, the cut
\ptpsq~$ < 2\times10^{-4}$~GeV$^2$/$c^2$ is applied.

The time of the \pipic\ decay is determined from hits
in the scintillator hodoscope associated with the tracks
and has a resolution of \about 150~ps.
The events with insufficient information to determine the decay time
accurately are discarded.
This inefficiency is 0.1\% and is measured to be equal for \KS\ and \KL.
 
\subsection{\boldmath \ks\ tagging}
\label{sec:tagging}

\begin{figure}[h!]
\vspace{-1.cm}
\begin{center}
\mbox{\epsfig{file=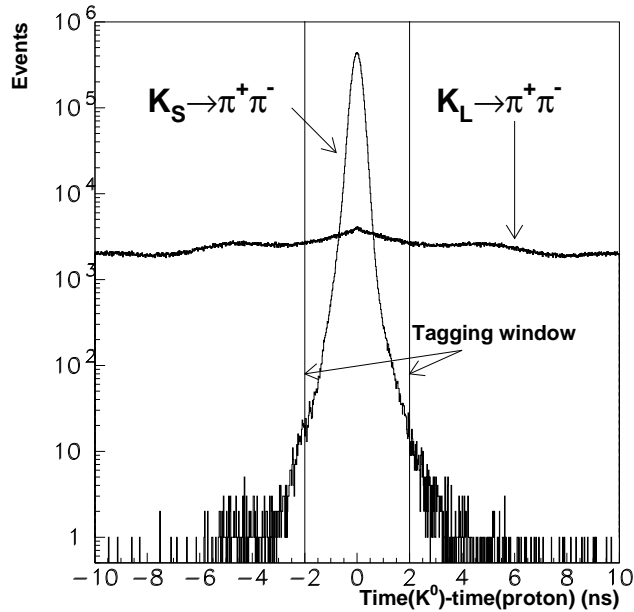,height=9cm,width=9cm}}
\end{center}
\rm
\caption{Time coincidence for \ks\ and  \kl\ \pipic\ decays, 
identified by their reconstructed vertex.}
\label{tagdis}
\end{figure}

A decay is labelled \ks\ if a coincidence is found (within a $\pm$2~ns interval)
between its event time and a proton time measured by the Tagger. 
Figure~\ref{tagdis} shows the time distributions 
for \ks\ and \kl\ decays to \pipic\, 
which have been identified as such by their vertex positions in the vertical 
direction. 
A similar procedure is not possible for \pipin\ decays; 
therefore  
tagging provides the only way to distinguish \ks\ from \kl.
The selection 
of \ks\ and \kl\ samples is done by means of tagging for
both the \pipic\ and \pipin\ modes, so that systematic effects are mostly
symmetric.
\par
The probability that a \ks\ decay is assigned to the \kl\ beam,
due to coincidence inefficiencies, is denoted by \asl.
This tagging inefficiency is directly measured in the \pipic\ 
mode, using the vertical vertex position, and is $(1.12 \pm 0.03)\E{-4}$,
dominated by the Tagger inefficiency.
The inefficiency could be different for \pipic\ 
and \pipin\ decays, since different detectors are used.
The difference between the two modes is estimated using
a large sample of \ks\ and \kl\ decays into 2$\pi^{0}$ and 3$\pi^{0}$
where one of the photons converts into an electron-positron pair. 
The time from the LKr clusters can then be compared with the 
scintillator-hodoscope time from the two tracks.
The conclusion is that the \pipin\ and \pipic\ tagging inefficiencies agree 
within an uncertainty of $\pm0.5\E{-4}$, which corresponds to an 
uncertainty on R of $\pm3\E{-4}$.
The tagging inefficiency can also be measured directly in \pipin\ events with a
subsequent \pioeeg\ Dalitz decay, which allows the \ks\ 
to be identified from the two-track vertex.
However this method is statistically limited. 
Another cross-check is performed in special runs where only \ks\ are present;
it gives a tagging inefficiency in agreement with the result above.

\par
The probability that a \kl\ decay is identified as a \ks\ decay
due to an accidental coincidence between the event and a proton
is called accidental tagging and is denoted as \als. It is measured 
in the \pipic\ mode to be  $(8.115\pm0.010)\times10^{-2}$
\footnote{It was $(10.649\pm0.008)\times10^{-2}$ in the 1998-99 data sample.}.
The \als\ difference, $\Delta \alpha_{LS}$,
between the \pipin\ and \pipic\ modes is estimated 
by measuring the probability to find a proton within time windows 4~ns wide,
located before or after the event time in tagged \kl\ events 
(i.e. events with no 
proton in coincidence). Ten windows are chosen, centred at 5~ns intervals 
from the event time so as to follow the 200~MHz structure of the proton beam. 
The extrapolation from the side windows to the coincidence window 
is performed in the \pipic\ mode with vertex-selected \kl\ and in the 
\pipin\ mode using \threepio\ decays, which come almost entirely from
the $K_L$ beam 
(the very small contribution from the $K_S$ beam being subtracted).
The measured value of $\Delta\als = \als^{00} - \als^{\pm} = (+3.4\pm1.4)\E{-4}$ 
corresponds to a correction on R of $(+6.9\pm2.8)\E{-4}$.
The origin of this effect is discussed in section~\ref{sec:dals}.

\vspace{0.4cm}
\subsection{Definition of the decay region}
\label{sec:sample}

 The fiducial ranges in kaon energy $E_K$ and in proper time $\tau$
used to count events are chosen to be $70< E_K < 170$~GeV and
$0 < \tau < 3.5\ \tau_S$, where $\tau=0$ is defined at the position
of the AKS counter and $\tau_S$ is the $K_S$ mean lifetime. 
For $K_L$ events, the decay time cut is applied on reconstructed $\tau$,
while for $K_S$ events the cut at $\tau=0$ is applied
using the AKS to veto decays occurring upstream. The nominal
$\tau=0$ positions defined by the AKS differ by $21.0\pm0.5$~mm between
$\pi^+\pi^-$ and $\pi^0\pi^0$ decays.
The veto inefficiency is 0.36\% for $\pi^0\pi^0$ events and
0.22\% for $\pi^+\pi^-$ decays. Given the fractions of
decays occurring upstream of the AKS position (respectively 5.8\% and
4.0\%), the correction to be applied to the double ratio
is $(1.2\pm0.3)\times10^{-4}$.

 $K_S \rightarrow \pi\pi$ decays can be produced in both beams by
scattering of beam particles in the collimators and, 
in the case of the $K_S$ beam,
in the AKS counter.
To reduce this contamination, a cut
on the extrapolated kaon impact point at the
level of the LKr calorimeter is applied to all events.
For $\pi^0\pi^0$ decays, it is
defined as the energy-weighted average $x,y$ positions of the four showers
at the face of the LKr calorimeter. For $\pi^+\pi^-$ decays, it is the
momentum-weighted average position of the tracks measured
upstream of the spectrometer magnet and projected onto the face of the
LKr calorimeter. 
The extrapolated kaon impact point is required to be within 10~cm of the
intersecting beam axes. 
The radii of the $K_L$ and $K_S$ beam spots are respectively
3.6~cm and 4.6~cm, so effects related to resolution smearing
are negligible.

\subsection{Data quality selection}
\label{sec:quality}

 Data collected in the $\pi^+\pi^-$ mode are affected by an overflow
condition in the drift chambers which resets the front-end readout buffers 
when there are more than seven hits in a plane within a 100~ns time
interval. This occurs mostly when an accidental particle generates
an electromagnetic shower upstream of the spectrometer and sprays
the drift chambers with particles. To maintain the highest reconstruction
and trigger efficiencies, events used in the analysis
are required to 
have no overflows within $\pm$~312~ns of the event time.
To minimise possible effects of $K_L$/$K_S$ intensity variation and
to equalise the beam intensities seen by good events,
this requirement is applied to both $\pi^+\pi^-$ and
$\pi^0\pi^0$ decays. The resulting event loss is 11\% in the
2001 data sample. This is significantly smaller than the 20\% loss
observed in the 1998-99 data, due to the lower instantaneous
beam intensity and to smaller noise in the drift chambers.
 Other dead time conditions affecting the first and the second
level $\pi^+\pi^-$ trigger ($\approx$ 0.3\% \footnote{It was 1.6\%
in the 1998-99 data.})
are also recorded and applied in the
analysis to $\pi^0\pi^0$ decays.

 Because of large beam intensity variations at the very beginning of the
spill, 
data from the first 0.2~s are not used in the analysis. The corresponding
loss of events is $\approx$ 1\% and cancels in the
ratio between $\pi^0\pi^0$ and $\pi^+\pi^-$ decays.

\section{R corrections and systematic uncertainties}
\label{sec:correction}

 The data are divided into 20 bins in kaon energy, each 5 GeV wide.
The numbers of $K_S$ and $K_L$ candidates
are corrected for the mistagging
probabilities discussed previously. 
The total numbers of events
are $1.546\times10^6$ $K_L\rightarrow\pi^0\pi^0$,
$2.159\times10^6$ $K_S\rightarrow\pi^0\pi^0$,
$7.136\times10^6$ $K_L\rightarrow\pi^+\pi^-$ and
$9.605\times10^6$ $K_S\rightarrow\pi^+\pi^-$.

Corrections for trigger efficiencies, background subtractions
and residual acceptance differences between $K_L$ and $K_S$ are 
applied separately in each energy bin before computing the average of $R$.

\subsection{\boldmath Trigger efficiencies}
\label{sec:trgeff}

The $\pi^0\pi^0$ trigger efficiency is measured using a control
sample of events triggered by a scintillating fibre detector located
inside the LKr calorimeter. The efficiency is found to be $(99.901\pm0.015)\%$.
The small inefficiency is $K_S$-$K_L$ symmetric and no correction
to the double ratio need be applied.

The $\pi^+\pi^-$ trigger efficiency is $(98.697\pm0.017)\%$
\footnote{It was $(97.782\pm0.021)\%$ in 1998-99. The improvement
comes from the lower beam intensity and the better efficiency of the
drift chambers in 2001.}.
The difference between the $\pi^+\pi^-$ trigger efficiency for
$K_S$ and $K_L$ decays is computed in each energy bin. The
overall correction on the double ratio is $(5.2\pm3.6)\times10^{-4}$, where
the uncertainty is given by the statistics of the control
samples used to measure the efficiency. 

\subsection{\boldmath Backgrounds}
\label{sec:bkg}
\subsubsection{Background to the \pipin\ mode}
\label{sec:bkgn}

The background to the $K_L\rightarrow\pi^0\pi^0$ signal comes uniquely from
\klthreepio\ decays, while the $K_S$ mode is background free.
The \klthreepio\ background has a flat $\chi^2$ distribution.
To estimate this background, a control region is defined by
$36<\chisq<135$. The excess of $K_L$ candidates in this
region over a Monte Carlo expectation for $\pi^0\pi^0$ decays
is used to extrapolate
the background in the signal region.
The Monte Carlo includes the effect of non-Gaussian tails 
in the calorimeter resolution so as to reproduce
the observed distribution in the $K_S$ sample.

The background is subtracted from the \kltwopio\ sample 
in bins of kaon energy
and the resulting correction on the double ratio, 
taking into account the uncertainties in the non-Gaussian tails and
in the background extrapolation,
is $(-5.6 \pm 2.0)\E{-4}$.   

\subsubsection{Background to the \pipic\ mode}
\label{sec:bkgc}

The background from $\Lambda \rightarrow p\pi^-$ 
in the \kstwopic\ sample is negligible after the cut
on the track momentum asymmetry.
\par

The residual \kethree\ and \kmuthree\ backgrounds 
in the $K_L$ sample are estimated by defining
two control regions in the \mpp-\ptpsq\ plane.
The first region, 
$9.5<(\mpp-\mk)< 19.0$ MeV/$c^2$ and $300<\ptpsq< 2000$ MeV$^2$/$c^2$, 
is dominated by \kethree\ events, 
while the second, 
$-17.0<(\mpp-\mk)< -12.0$ MeV/$c^2$ and $300<\ptpsq< 500$ MeV$^2$/$c^2$, 
contains roughly equal numbers of \kethree\ and \kmuthree\ events. 

The background distributions in the control regions are modelled by a  
\kethree\ sample, selected with $\eop>0.95$,
and by a \kmuthree\ sample, obtained by reversing the muon veto requirement; 
the tails in the \kltwopic\ distribution
are estimated from the \ks\ sample. The result is then 
extrapolated to the signal region.

The overall \kethree\ background fraction is $10.5\E{-4}$, 
the \kmuthree\ background is  $4.0 \E{-4}$. 
The background subtraction is applied in bins of kaon energy and the
resulting correction on the double ratio is $(14.2 \pm 3.0)\E{-4}$, where 
the error has been estimated by changing the control regions 
and the modelling of the resolution tails.

Kaon decays to $\pi^+\pi^-\gamma$ have been shown to
have a negligible effect on R \cite{na48_99}.

\subsubsection{Collimator scattering}
\label{sec:colsca}

In the $K_S$ beam, the cut on the extrapolated kaon impact point
is stronger than the \ptpsq\ cut applied to
\pipic\ decays, and therefore the contribution of beam scattering
is removed symmetrically from both final states. 
On the contrary, in the $K_L$ beam, the \ptpsq\ cut which
is applied only in the \pipic\ mode is stronger and therefore the
small residual contribution from scattered events must be subtracted
from the \pipin\ sample.

The correction for this asymmetry is computed from reconstructed 
$\kltwopic$ candidates with an inverted \ptpsq\ cut. 
The scattered events are extracted from the peak
at the kaon mass in the \mpp\ invariant mass distribution.
The correction to R is applied in bins of energy and 
it amounts to $(-8.8 \pm 2.0)\E{-4}$.

\subsection{\boldmath Acceptance}
\label{sec:accep}

The \ks\ and \kl\ acceptances are made very similar in both modes 
by weighting \kl\ events according to their proper decay time.
The weighting factor takes into account the small interference term.
A small difference in acceptances remains, 
related to the differences in \ks\ and \kl\ beam sizes and directions. 
This residual correction is computed using a large-statistics Monte Carlo simulation 
(4$\E{8}$ generated kaon decays per mode). 
The largest contribution to the correction comes from 
the difference between the \ks\ and \kl\ beams 
near the beam axes in the spectrometer for \pipic\ decays. The acceptance 
correction related to the \pipin\ mode is small.

The systematic uncertainty on the acceptance correction is evaluated 
by varying the $K_S$ beam halo, the beam positions and shapes,
and the drift-chamber inefficiencies. The resulting
systematic uncertainty is $\pm3.0\times10^{-4}$.
A detailed comparison between a fast simulation and a GEANT~\cite{geant} based
simulation of the spectrometer was performed on the 1998-99 sample.
This resulted in an additional systematic uncertainty of
$\pm2.3\times10^{-4}$.
The final correction to \R\ for
the acceptance is: 
$\DR(\mbox{acceptance}) = (+21.9 \pm 3.5(\mbox{MC} \stat) \pm 4.0(\syst))\E{-4}$.

\subsection{\boldmath Energy and distance scales}
\label{sec:scale}

 The determinations of the kaon energy, the decay vertex and the proper
time in the $\pi^0\pi^0$ mode rely on measurements of the photon
energies and positions with the calorimeter.

 The absolute energy scale is adjusted using $K_S \rightarrow
\pi^0\pi^0$ decays. 
The energy scale is set such that the average value of the reconstructed
decay position in a range centred around the anticounter matches the value
found in a Monte Carlo simulation. 
This measurement of the energy scale is checked using data taken
during special runs (so-called $\eta$ runs)
with a $\pi^-$ beam striking two thin targets
located near the beginning and the end of the fiducial decay region,
producing $\pi^0$ and $\eta$ with known decay positions.
From two-photon decays of $\pi^0$ and $\eta$, the reconstructed
vertex position can be computed using the $\pi^0$ or the $\eta$ mass
value (the $\eta$ mass value is taken from \cite{eta_mass}), and
compared to the nominal target positions. Continuum production
of $2\pi^0$ events is also used, with the advantage that the
final state is very similar to the one of kaon decays. The two
targets give energy scales consistent to better than $10^{-4}$.
The uncertainty on the overall energy scale is estimated from
these comparisons to be $\pm 3\times10^{-4}$. The corresponding
uncertainty on the double ratio is $\pm 2\times10^{-4}$.

 Non-linearities in the energy response are studied using K$_{e3}$ decays,
where the electron energy measured in the calorimeter can be compared to the
momentum measured in the spectrometer, and using data from the
$\eta$ runs. Parameterising the deviations from linearity as
$\Delta E/E = \alpha/E +\beta\cdot E$, $\alpha$ is constrained
to $\pm$10 MeV and $\beta$ is bound to be in the range 
$\pm2\times10^{-5}$~GeV$^{-1}$. Taking also into account larger
deviations from linearity observed in the region $E_{\gamma}<6$~GeV,
the resulting uncertainty on the double ratio is $\pm3.8\times10^{-4}$.

 The uniformity of the calorimeter response over its surface is optimised
using K$_{e3}$ decays and checked using $\pi^0$ decays from the
$\eta$ runs. Bias on the double ratio can arise from a dependence
of the energy response on the photon impact radius $r$.
Parameterising this effect as $\Delta E/E = \gamma\cdot r$,
$\gamma$ can be bound to be in the range $\pm10^{-3}$~m$^{-1}$.
In the region close to the beam tube,
residual variations are smaller than 0.2\%. The systematic uncertainty
on the double ratio from these effects is $\pm1.6\times10^{-4}$.

 Uncertainties in the correction of energy leakage from one
cluster to another can lead to an apparent non-linearity
and bias the double ratio. The correction used in the data is based
on the transverse shower-profile measured during special runs 
in which single monochromatic electrons
were sent to the calorimeter. The uncertainty in the shower profile
is taken to be the difference between this measurement and the prediction
of the GEANT Monte Carlo simulation. The resulting uncertainty
on the double ratio is $\pm1.1\times10^{-4}$.

 The measurements of photon positions and the transverse size-scale
of the calorimeter are adjusted and checked using K$_{e3}$ decays,
comparing the reconstructed cluster position with the electron track
impact point extrapolated to the calorimeter. The associated
uncertainty on the double ratio is $\pm1.6\times10^{-4}$.

 In the computation of the decay vertex position, the photon positions
must be extrapolated to the longitudinal position of the maximum
of the shower to account correctly for deviations of the photon directions
from the projectivity of the calorimeter.
Comparing data and Monte Carlo simulation
in $K_{e3}$ decays, the uncertainty on this
position is $\pm2$~cm. The resulting uncertainty on the double ratio
is $\pm1.6\times10^{-4}$.

 Finally, the effect of non-Gaussian tails in the energy response is
minimised by the choice of the procedure used to adjust the overall
energy scale. Residual effects are investigated by applying to the
Monte Carlo samples a parameterisation of non-Gaussian tails in the
energy response (arising mostly from photo-production of hadrons
early in the electromagnetic shower) derived from $K_{e3}$ and
$\eta$ data. No bias on the double ratio is observed within the
Monte Carlo statistical error which is $\pm1.0\times10^{-4}$.

 Adding all the above uncertainties in quadrature, the
total systematic error 
on the double ratio from the measurements of the photon
energies and positions is found to be $\pm5.3\times10^{-4}$.

 For \pipic\ decays, the vertex position is measured from
the reconstructed tracks and is completely determined by the
detector geometry. As a check, the reconstructed anticounter
position can be measured in $K_S \rightarrow \pi^+\pi^-$ decays.
The value obtained agrees with the nominal position to better 
than 1~cm \footnote{The difference was 2~cm in the 1998-99 data.}.
Uncertainties in the geometry of the detector are at the
level of 2~mm for the distance between the first two drift chambers,
and 20~$\mu$m/m for their relative transverse scale. This corresponds
to possible deviations of $\approx$ $\pm2$~cm on the
reconstructed AKS position. The corresponding uncertainty on the
double ratio measurement is $\pm2.0\times10^{-4}$. The asymmetry
in the $K_S$ and $K_L$ event losses, which could arise
from the effect of non-Gaussian tails in the \ptpsq\ resolution,
is smaller than $2.0\times10^{-4}$.
 The overall uncertainty on the double ratio from the
reconstruction of $\pi^+\pi^-$ decays is therefore $\pm 2.8\times10^{-4}$.

\section{Intensity effects}
\label{sec:intensity}

\subsection{\boldmath Uncertainty on R due to accidental effects}
\label{sec:accidental}

 Most of the accidental activity in the detector
is related to kaon decays in
the high-intensity $K_L$ beam. The
overlap of extra particles with a good event may
result in the loss of the event in the reconstruction or the
selection \footnote{The losses in the trigger are already
accounted for in the measurement of the trigger efficiency.}, 
depending on the time
and space separation of the activity in the detector. 
This effect is minimised by
the simultaneous collection of data in the four channels and
by the fact that $K_S$ and weighted $K_L$ decays illuminate
the detector in a similar way.
 The possible residual effect on the double ratio can be separated
into two components: intensity variations between the two beams coupled
to different, intensity-dependent,
event losses in the $\pi^+\pi^-$ and $\pi^0\pi^0$ modes
(intensity difference), and
a residual difference in the illumination between $K_S$ and
$K_L$ decays coupled to a variation of the event loss with the
impact points of the 
$K^0$ decay products (illumination difference).

\subsubsection{\boldmath Intensity difference}
If the losses depend, as expected,
linearly on the $K_L$ beam intensity, the intensity difference
effect is given by:
\begin{equation}
\Delta R = \Delta\lambda \times \Delta I/I
\end{equation}
where $\Delta\lambda$ is the difference between the mean
losses in $\pi^+\pi^-$ and $\pi^0\pi^0$, and $\Delta I/I$ is
the difference in the mean $K_L$ beam intensity as seen by
$K_L$ and $K_S$ events.

\begin{figure*}[htb]
 \vspace{-1cm}
 \begin{center}
 \mbox{\epsfig{file=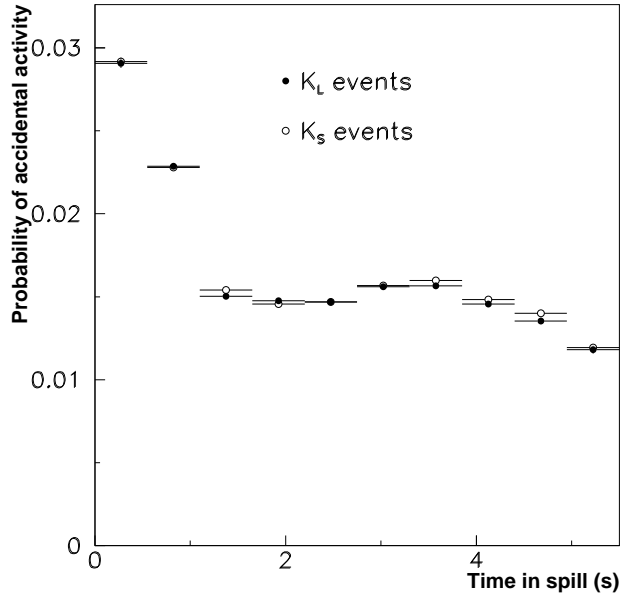,height=9cm,width=9cm}}
 \caption{Probability of accidental activity in the LKr in the $\approx$~150~ns
 readout window, as a function of the time during the spill, for $K_S$ and $K_L$
 decays to $\pi^+\pi^-$.}
  \label{fig:acci}
 \end{center}
\end{figure*}

  The accidental rate can be measured directly from the activity
in the detector within the readout time-window before each event.
Comparing the rate of out-of-time LKr clusters and
out-of-time tracks in good $K_S,K_L \rightarrow \pi^+\pi^-$ decays,
$\Delta I/I$ is found to be respectively ($+0.4\pm0.4$)\% and
($+0.6\pm0.3$)\%, where the quoted uncertainties are only statistical.
For this measurement $K_S$ and $K_L$ are identified by the
decay vertex position to avoid the correlation between mistagging probability
and beam intensity. The bias on
$\Delta I/I$ from the fact that this measurement uses good
reconstructed decays is a negligible second-order effect.
The measurement of accidental activity is illustrated
in Figure~\ref{fig:acci}.

The $K_L$ beam
intensity for each event can also be estimated using
the information from the $K_L$ beam monitor, integrating the
intensity over a 200~ns time window. The difference between the
average intensities as seen by $K_S$ and $K_L$ decays is found
to be ($-0.08\pm0.04$)\%, where the quoted uncertainty is only
statistical. The systematic uncertainty on this measurement is estimated
to be less than 1\%.
Another method to study the correlation of the two beams, which
does not rely on the use of good events, consists of a direct
computation of the correlation between the $K_L$ and $K_S$ beam
monitor counts, using events taken uniformly in time. To
avoid statistical fluctuations in the rate measurements, this is
done using a 15~$\mu$s integration time of the beam monitors. 
All known $K_L$/$K_S$ variations take place on a much longer time scale.
This method confirms that $\Delta I/I$ is consistent with 0 within
1\%. The final estimate for $\Delta I/I$ is $(0\pm1)\%$.

\par 
 The beam-induced event losses can be evaluated by overlaying
data with beam monitor (BM)
triggers taken in proportion to the beam intensity, 
which reproduce the ambient activity as seen by the detectors.
Detector information from the original events and the BM triggers
are superimposed at the raw-data level and the overlayed events undergo
the standard reconstruction and analysis procedure.
Losses and gains from migration of events around the cuts are
taken into account. When
estimating the event losses related to accidental activity, the losses
related to doubling the noise effect inherent in this procedure must
be removed.
The net event losses induced by
the accidental activity are at the level of 1 to 2\%. 
The dominant source of event loss is found to be the appearance
of the drift chamber overflow condition
in the overlayed event.
This loss is higher for $\pi^+\pi^-$ decays than for
$\pi^0\pi^0$ because of the presence 
of at least two drift chamber hits per plane
for the original $\pi^+\pi^-$ decay.
From this procedure, $\Delta\lambda$ is found to be 1.0\%.
The overlay procedure can also be applied starting from a Monte Carlo
sample of original $\pi^+\pi^-$ and $\pi^0\pi^0$ decays. In this
case, $\Delta\lambda$ is found to be 0.65\%. The statistical
uncertainties on $\Delta\lambda$ from the overlay samples
are less than 0.1\%.
The difference between the two estimates is attributed
to the lower hit-multiplicity in the Monte Carlo samples compared to
the data.

 $\Delta\lambda$ can also be estimated directly from the
data, by comparing the ratio of \pipic\ and \pipin\ events 
obtained in
the normal $K_L+K_S$ beam runs and in pure $K_S$ runs, in which
the accidental activity is one order of magnitude smaller.  This
leads to $\Delta\lambda = (0.9\pm0.6)\%$. Similarly,
$\Delta\lambda$ can be checked by dividing the data into bins of
$K_L$ beam intensity and looking at the variations in the ratio
of \pipic\ and \pipin\ events. This leads to an estimate in agreement
with the values above. In conclusion, from this study
the final estimate of $\Delta\lambda$ is $(1.0\pm0.5)\%$.
This value is typically 30\% lower than the estimate derived for the
1998-99 data-taking period, as expected from the lower beam intensity
in 2001.

\begin{figure*}[htb]
 \vspace{-1cm}
 \begin{center}
 \mbox{\epsfig{file=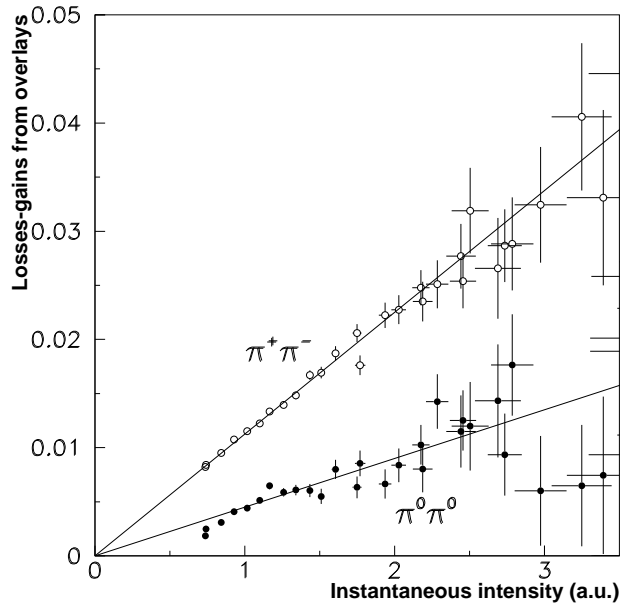,height=9cm,width=9cm}}
 \caption{Beam-induced net event losses as a function of instantaneous intensity
 in arbitrary units.
 Lines are drawn to guide the eyes.}
  \label{fig:loss_vs_i}
 \end{center}
\end{figure*}

 Finally, the linearity of the losses with intensity can be
checked by looking at the losses as a function of the beam intensity
given by the BM triggers. 
 Figure~\ref{fig:loss_vs_i} shows the accidental event losses in
the $\pi^+\pi^-$ and $\pi^0\pi^0$ modes as a function of the beam
intensity from the overlay procedure applied to Monte Carlo events.

 Taking into account all the above results, the estimate of the uncertainty
on the double ratio related to differences in intensity-dependent
losses is $\pm1.1\times10^{-4}$
\footnote{For the 1998-99 sample, this uncertainty was estimated to be
$\pm3.0\times10^{-4}$.}.

\subsubsection{\boldmath Illumination difference}

The illumination-difference
effect has been estimated from the overlay samples, computing
separately the losses for $K_S$ and $K_L$ events.
This computation has been performed using both data and Monte Carlo
original events. In the first case,
the value obtained is (+0.9$\pm$3.5)$\times$10$^{-4}$, in the
second it is (+1.4$\pm$2.8)$\times$10$^{-4}$, where the quoted
uncertainties are the statistical errors from the overlay samples.
As expected, there is no evidence of a significant effect, and we
use as uncertainty on the double ratio $\pm3.0\times10^{-4}$.

\subsubsection{\boldmath Overall uncertainty on R and cross-check}

 Combining the two above uncertainties in quadrature, the
total uncertainty on R from accidental effects is
$\pm3.1\times10^{-4}$. This uncertainty is dominated by the
statistical error of the overlay procedure.

 The overlay method can also be used to estimate the combination of
the two effects. For this, $K_S$ events are overlayed
only with BM triggers from the $K_S$ monitor and $K_L$ events
only with BM triggers from the $K_L$ monitor. This method relies
on the accuracy of the BM triggers to estimate
correctly the $K_L$-$K_S$ intensity difference. 
From this method, the overall accidental effect on the double
ratio is found to be $(4.7\pm4.9)\times10^{-4}$, where the quoted
error is the statistical uncertainty from the overlay sample.
The systematic uncertainty from the accuracy of the BM triggers
is expected to be $<$ 2$\times$10$^{-4}$. From this
cross-check, there is no evidence of 
any unexpected effect on the double ratio.

\subsubsection{{\boldmath In-time activity from the} $K_S$ {\boldmath target}}

 The techniques above do not take into account any additional detector
activity in the $K_S$ beam generated by the same proton which produced
the $K_S$ event. Studies of this background, mostly
searching in the LKr calorimeter for additional clusters in $2\pi^0$ events
from pure $K_S$ beam runs, allow an upper bound on the
effect on R of $1\times10^{-4}$ to be set.

\subsection{\boldmath Origin of $\Delta\als$}
\label{sec:dals}

 The accidental tagging probability \als\ depends only on the proton
beam intensity seen by the Tagger and, consequently, it is to first
order the same for \pipin\ and \pipic . However, because the event
selection is more sensitive to accidentals for \pipic\ events, we
expect a difference of the measured \als\ for the \pipin\ and \pipic\
samples due to beam-intensity variations with time.
 A quantitative understanding of the effect of accidental activity on 
selected events
can be reached by studying the BM trigger overlays.

 A value $\Delta\als=(+3.5\pm0.4)\times10^{-4}$ is expected
from the overlay of data,
where the error is only statistical. Using loss and gain probabilities from
the overlay Monte Carlo samples instead of data we find
$\Delta\als=(+2.0\pm0.4)\times10^{-4}$.

\begin{figure}[h!]
\vspace{-1.cm}
\begin{center}
\mbox{\epsfig{file=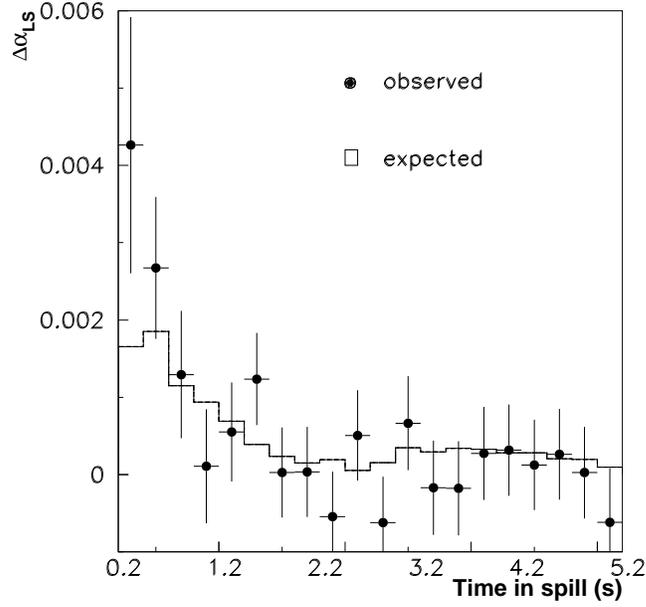,height=9cm,width=9cm}}
\end{center}
\rm
\caption{Measured compared
to predicted values of  $\Delta\als$ as a function of the time
during the spill}
\label{randals}
\end{figure}

 In Figure~\ref{randals} the variation of $\Delta\als$ 
within the spill is shown and compared with the overlay computation. Most of
the difference in accidental tagging between the \pipin\ and \pipic\
modes comes from the beginning of the spill where the instantaneous
intensity is higher and the beam intensity variations more significant.

 Another source of event losses is the inefficiency of the \pipic\ trigger.
The intensity-dependent part is studied separately and its
effect on $\Delta\als$ is estimated to be $(0.4\pm0.2)\times10^{-4}$.

 The observed $\Delta\als$ value of
$(+3.4\pm1.4)\times10^{-4}$ is therefore well reproduced both
qualitatively and quantitatively.
 
\section{Result}

The effect on the result of the corrections described above,
and the various sources of systematic uncertainties
are summarised in Table~\ref{tab:result}.

\begin{table}[!htb]
\begin{center}
\caption{Corrections and systematic uncertainties on the double
ratio R (2001 data)}
\label{tab:result}
\begin{tabular}{|l|rrr|} \hline
&\multicolumn{3}{c|}{in 10$^{-4}$} \\ \hline
\pipic\ trigger inefficiency                    &  $+5.2$       & $\pm$ 3.6  & (stat) \\
AKS inefficiency                                &  $+1.2$       & $\pm$ 0.3  & \\
Reconstruction \begin{tabular}{@{}l} of \pipin\ \\ of \pipic\ \end{tabular} &
\begin{tabular}{r@{}}    ---   \\  ---   \end{tabular} &
\begin{tabular}{r@{}}   $\pm$ 5.3 \\  $\pm$ 2.8  \end{tabular} & \\
Background \begin{tabular}{@{}l} to \pipin\ \\ to \pipic\ \end{tabular} &
\begin{tabular}{r@{}} $-5.6$ \\ $+14.2$ \end{tabular} &
\begin{tabular}{r@{}} $\pm$ 2.0  \\ $\pm$ 3.0  \end{tabular} & \\
Beam scattering                                 &  $-8.8$       & $\pm$ 2.0  & \\
Accidental tagging                              &  $+6.9$       & $\pm$ 2.8  & (stat)  \\
Tagging inefficiency                            &  ---          & $\pm$ 3.0  & \\
Acceptance \begin{tabular}{@{}l} statistical \\ systematic \end{tabular}
                              & $+21.9$ &
     \begin{tabular}{r@{}} $\pm$ 3.5 \\ $\pm$ 4.0 \end{tabular}  & 
 \begin{tabular}{r@{}} (stat) \\ \ \end{tabular} \\
Accidental activity \begin{tabular}{@{}l} intensity difference \\
                                          illumination difference \end{tabular}
                            &  ---    &   \begin{tabular}{r@{}} $\pm$ 1.1 \\
                                          $\pm$ 3.0 \end{tabular}  &
\begin{tabular}{r@{}} \\  (stat) \end{tabular} \\
$K_S$ in time activity      & ---     & $\pm$ 1.0 & \\
\hline
Total                                           & $+35.0$       & $\pm$ 11.0 & \\
\hline
\end{tabular}
\end{center}
\end{table}

\begin{figure*}[htb]
 \vspace{-1cm}
 \begin{center}
 \mbox{\epsfig{file=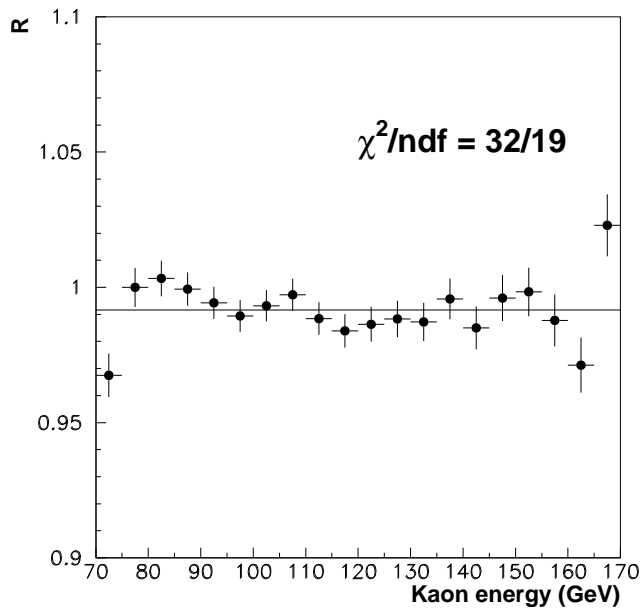,height=9cm,width=9cm}}
 \caption{Measured double ratio R in kaon energy bins.}
 \label{fig:r_vs_e}
 \end{center}
\end{figure*}
 
Figure~\ref{fig:r_vs_e} shows the measured double ratio after corrections
as a function of the kaon energy.

The final result for the double ratio from the 2001 data set
is $R=0.99181\pm0.00147\pm0.00110$, where the first error
is the statistical error from the 2$\pi$ samples, and
the second is systematic. Out of this systematic
uncertainty, $\pm0.00065$ is due to the finite
statistics of the control samples used to study the systematic
effects.

Many cross-checks of the stability of the result have been performed,
by varying
some of the selection cuts and by searching for a dependence of the
result on several variables,
such as the beam intensity, the time
during the spill, and the data-taking period. No significant variation in the
result is observed.
An analysis adopting a different scheme for data
compaction, filtering, selection and correction procedures
was also performed in addition to the one presented here. Its
result fully confirms the above measurement.

The corresponding value of the direct CP-violation parameter
Re($\epsilon'/\epsilon)$ from Formula \ref{eq:doub} is: 
\begin{center}
Re($\epsilon'/\epsilon)$ = $(13.7\pm2.5\pm1.1\pm1.5)\times10^{-4}$ ,\\
\end{center}
where the first uncertainty is the pure statistical error from the
2$\pi$ samples, the second is the systematic error coming from
the statistics of the control samples, and the third is the contribution
of the other systematic uncertainties.
Combining the errors in quadrature, the result is:
\begin{center}
Re($\epsilon'/\epsilon)$ = $(13.7\pm3.1)\times10^{-4}$. \\
\end{center}

This result is in good agreement with the published value
from the 1997-98-99 data:
Re($\epsilon'/\epsilon)$ = $(15.3\pm2.6)\times10^{-4}$.

The comparison of the present and earlier results is particularly
significant~since~they were~obtained~from data
taken at different average beam intensities.
The~correlated~systematic uncertainty between the
two results is estimated to be $\pm1.4\times10^{-4}$.
Taking this~correlation into account, the combined, final result on
Re($\epsilon'/\epsilon)$ from the NA48 experiment~is:
\begin{center}
Re($\epsilon'/\epsilon)$ = $(14.7\pm1.4\pm0.9\pm1.5)\times10^{-4}$, \\
\end{center}
or, with combined errors:
\begin{center}
Re($\epsilon'/\epsilon)$ = $(14.7\pm2.2)\times10^{-4}$. \\
\end{center}

\section*{Acknowledgements}

It is a pleasure to thank the technical staff of the participating
laboratories and universities for their efforts in the design and
construction of the apparatus, in the operation of the experiment, and
in the processing of the data.

\end{document}